\documentclass[english,showpacs,twocolumn]{revtex4-2}

\usepackage[T1]{fontenc}
\usepackage[utf8]{inputenc} 

\usepackage{xcolor}
\usepackage{graphics}
\usepackage{graphicx}
\graphicspath{{Figures/}}
\usepackage{epsfig}
\usepackage{bm}
\usepackage{epstopdf}
\usepackage{hyperref}
\usepackage{tikz}
\usepackage{xcolor}

\usepackage{mathptmx}
\usepackage{amsmath}
\usepackage{amssymb}
\usepackage{mathtools}
\usepackage{float}
\usepackage{bm}

\usepackage{dcolumn}

\AtBeginDocument{%
    \newwrite\bibnotes
    \def\bibnotesext{Notes.bib}
    \immediate\openout\bibnotes=\jobname\bibnotesext
    \immediate\write\bibnotes{@CONTROL{REVTEX42Control}}
    \immediate\write\bibnotes{@CONTROL{%
    apsrev42Control,author="08",editor="1",pages="1",title="0",year="1"}}
     \if@filesw
     \immediate\write\@auxout{\string\citation{apsrev42Control}}%
    \fi
}

\makeatletter
\def\p@figure{\color{blue}}
\def\p@equation{\color{blue}}
\makeatother

\begin{document}

\title{Aggregate morphing of self-aligining soft active disks in semi-confined geometry}
\author{Anshika Chugh$^{1,2}$}
\email{anshikachugh93@gmail.com}
\author{Soumen De Karmakar$^{3}$}
\email{soumendekarmakar@gmail.com}
\author{Rajaraman Ganesh$^{1,2}$}
\email{ganesh@ipr.res.in}
\affiliation{$^{1}$Institute for Plasma Research, Bhat, Gandhinagar 382428, India}
\affiliation{$^{2}$ Homi Bhabha National Institute, Training School Complex, Anushaktinagar, Mumbai 400094, India}
\affiliation{$^{3}$ Institute for Theoretical Physics IV, University of Stuttgart, Heisenbergstr. 3, Stuttgart 70569, Germany}
\date{\today}

\begin{abstract}
We study the dependence of alignment and confinement on the aggregate morphology of self-aligning soft disks in a planer box geometry confined along $y$ direction. We show that the wall accumulation of aggregates becomes non-uniform upon increase in alignment strength and decrease in box width. The height of these structures is found to be a non-monotonic function of alignment strength. Additionally, we identify two distinct categories of wall aggregates: layered and non-layered structures each exhibiting distinct local structural properties. For non-layered structures, local properties stay nearly constant as we move away from the boundary, while for layered structures, they increase with distance from the boundary. Our analysis shows that active pressure difference is a useful indicator for different aggregate morphologies and the peaks in the pressure curve are indicative of the average and minimum height of the structure.
\end{abstract}

\maketitle

\section{\label{sec:Intro}Introduction}
Motility Induced Phase Separation (MIPS) \cite{cates2015motility} is an interesting collective phenomenon that is unique to active matter systems. Due to particles' self-propulsion, uniformly distributed particles tends to spontaneously phase separate into dense and dilute regions. However, this phase separation requires sufficiently high motility and critical packing fraction. Interestingly, under certain alignment \cite{vicsek1995novel, kudrolli2008swarming, de2022reentrant} rules between particles, MIPS can been observed even at low packing fraction. Studies on MIPS has a rich literature encompassing investigations into the dynamics within both non-inertial \cite{caporusso2020motility, bialke2015active} and inertial limits \cite{lowen2020inertial, omar2023tuning, de2022motility}, with and without external alignment among particles \cite{vicsek1995novel, caprini2020spontaneous}, and there have been few recent studies with soft particles \cite{de2022motility, hopkins2023motility, sanoria2021influence}. Majority of these studies focus on the bulk behavior and properties in unbounded or doubly-periodic systems. The structural and dynamical properties of clusters or aggregates generated in MIPS have been intensively studied in bulk systems. These studies have yielded important insights into the emergent properties \cite{vicsek1995novel, kudrolli2008swarming, caporusso2020motility, caprini2020spontaneous} of active particles. However, studying collective phenomenon in confined and crowded environments is crucial, as various active entities are naturally found in confined spaces such as soil (porous media) \cite{iyer2023dynamics}, narrow channels (blood vessels) \cite{ao2014active, law2022microrobotic} and biofilms \cite{mazza2016physics}.

Understanding how active particles behave in confined spaces has potential applications in sorting and drug delivery \cite{abdelmohsen2014micro, hulme2008using}. Active particles do not immediately bounce off the boundaries due to their persistence. Consequently, the particles are accumulated at the boundaries that has been observed in experiments \cite{di2010bacterial, kumar2019trapping, galajda2007wall} and in numerical simulations \cite{wan2008rectification, fily2014dynamics, kaiser2012capture}. It gives insights on the pattern formation of bacteria in natural settings \cite{mazza2016physics}. Moreover, boundary shape strongly influences the spatial distribution of the particles when the size of the confined space is smaller than the persistence length \cite{fily2014dynamics}. Other studies demonstrate the effect of particle-boundary interactions \cite{reichhardt2011active}, and particle-particle interactions \cite{yang2014aggregation} on the accumulation of particles at the boundary which leads to rectification effects and spontaneous segregation of the particles. A recent study on MIPS in confined environments include that in porous environment such as porous walls, where the clusters undergo a morphological transition from a uniform accumulation at walls to accumulation at specific locations on the wall \cite{das2020aggregate, das2020morphological}. Considerable efforts have been made to understand the dynamics of active particles in confined spaces by investigating the effects of boundary shape \cite{fily2014dynamics}, particle-wall interaction \cite{reichhardt2011active}, the influence of inter-particle interaction \cite{yang2014aggregation}, as well as by changing wall property such as making the wall porous \cite{das2020morphological}. Little attention has been given to considering particle alignment as an alternative approach to influence the dynamics of wall accumulation. This study investigates how particle alignment affects the dynamics of active particles in confined spaces.
\begin{figure}
  \includegraphics[width = 0.98\linewidth, height=0.47\linewidth]{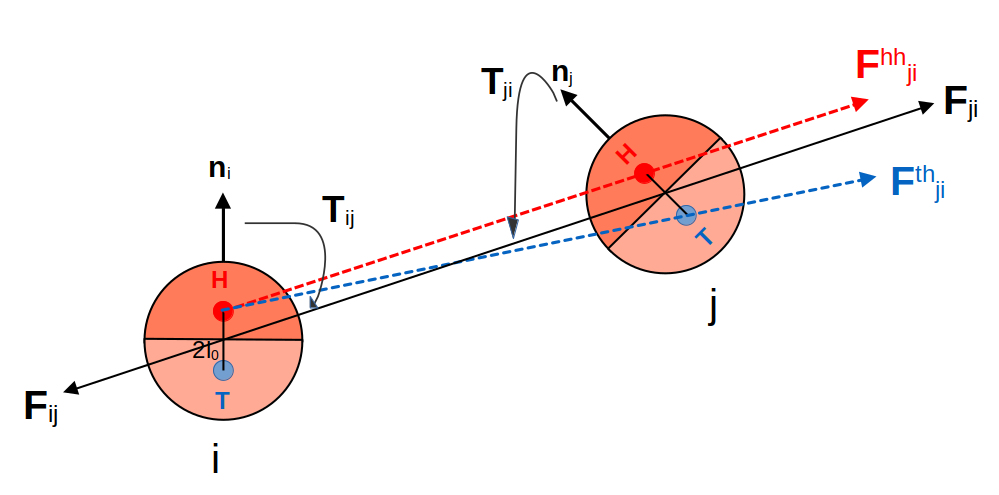}
  \caption{\label{fig:fig0} Schematic of alignment model \cite{de2022reentrant}. The head
and tail force centers, denoted by the red and blue dots, respectively, are located $2 l_0$ distance
apart. The head force center of the $i$th disk interacts with the head (indicated by a red arrow)
and the tail (indicated by a blue arrow) force centers of the $j$th disc, giving rise to a resultant
force represented by a black arrow. This results in nonreciprocal torque ($T_{ij} /T_{ji}$) which aligns the self-propulsion axis ($n_i /n_j$) of both of the disks along the inter-particle separation and towards each other.}
\end{figure}

A recent study on self-aligned particles introduces a novel model for reorientation or alignment. This model involves the non-reciprocal alignment of self-propulsion directions of the particles along their inter-particle distance and towards or away from the other particles \cite{de2022reentrant, zhang2021active}. Interestingly, this model exhibits both the collective phenomenon of MIPS \cite{de2022reentrant} and flocking \cite{thesisSoumen} depending upon the turning direction of particles towards or away from each other as they approach. In this study, this same non-reciprocal alignment model is considered where two active (disk) particles turn their self-propulsion direction towards each other and along the line of separation between these particles as they approach. This alignment model, depicted in Fig. \ref{fig:fig0} considers different strength of interaction between the the two halves of the particles, that naturally gives rise to torque that aligns the particles towards or away from each other based on the whether the repulsive force between the tails of the two particles exceeds the repulsive force between their heads or vice versa.

It is important to note that MIPS is observed at densities well below the critical threshold, provided a certain alignment rule is in place which orients the particles towards each other \cite{de2022reentrant}. Therefore, this study considers low density where conventional MIPS using non-aligning particles is not observed. In this low density scenario, inertia of the particle becomes important \cite{thesisSoumen}. Additionally, this study focuses on soft particles that substantially deform upon collision, thus allowing for its deformation \cite{de2022motility, thesisSoumen, sanoria2021influence}. This is relevant in the context of biological active species which are soft in comparison to the their synthetic counterparts such as self-propelled colloids. For doubly periodic systems, Ref. \cite{de2022motility} reports a comprehensive parameter analysis for the role of inertia, interaction strength, particle's softness and motility (Peclet number) on MIPS behavior and the related aggregate properties. The study reveals that increasing inertia and softness suppresses MIPS, while increasing motility or Peclet number promotes MIPS, enabling its occurrence even at relatively low values of particle's softness. Additionally, as interaction strength increases, MIPS exhibits re-entrant behavior for a given softness, inertia and motility. In the present study, we consider soft disks with finite inertia in a low density system such that in the absence of alignment between the particle, MIPS is absent. 

The following sections provides the model, results and the related discussions.

\section{\label{sec:Model}Model}
\begin{figure}
  \includegraphics[width = 0.98\linewidth, height=0.6\linewidth]{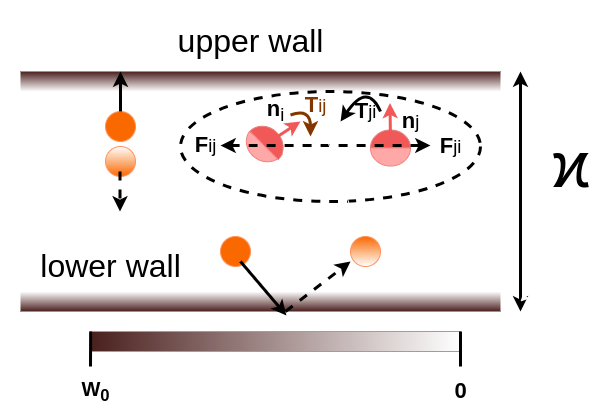}
  \caption{\label{fig:fig1} Illustrative diagram of particle-wall interaction and particle-particle interaction. Color bar denotes the strength of potential at the wall. In the hard wall limit, only the particles very close to the wall feel high repulsive force and are reflected from the wall boundary. Solid and dashed arrow show the particle trajectory before and after collision respectively. Particle-particle interaction is shown in the dashed circle, where the head and tail of the one disk separately interacts with sub-hemispheres of the other disk and the torque resulting from these interactions turns the self-propulsion of the disks towards each other. $\varkappa$ denotes the channel width.}
\end{figure}
\begin{figure*}[t]
  \includegraphics[width = 0.35\linewidth, height=0.25\linewidth]{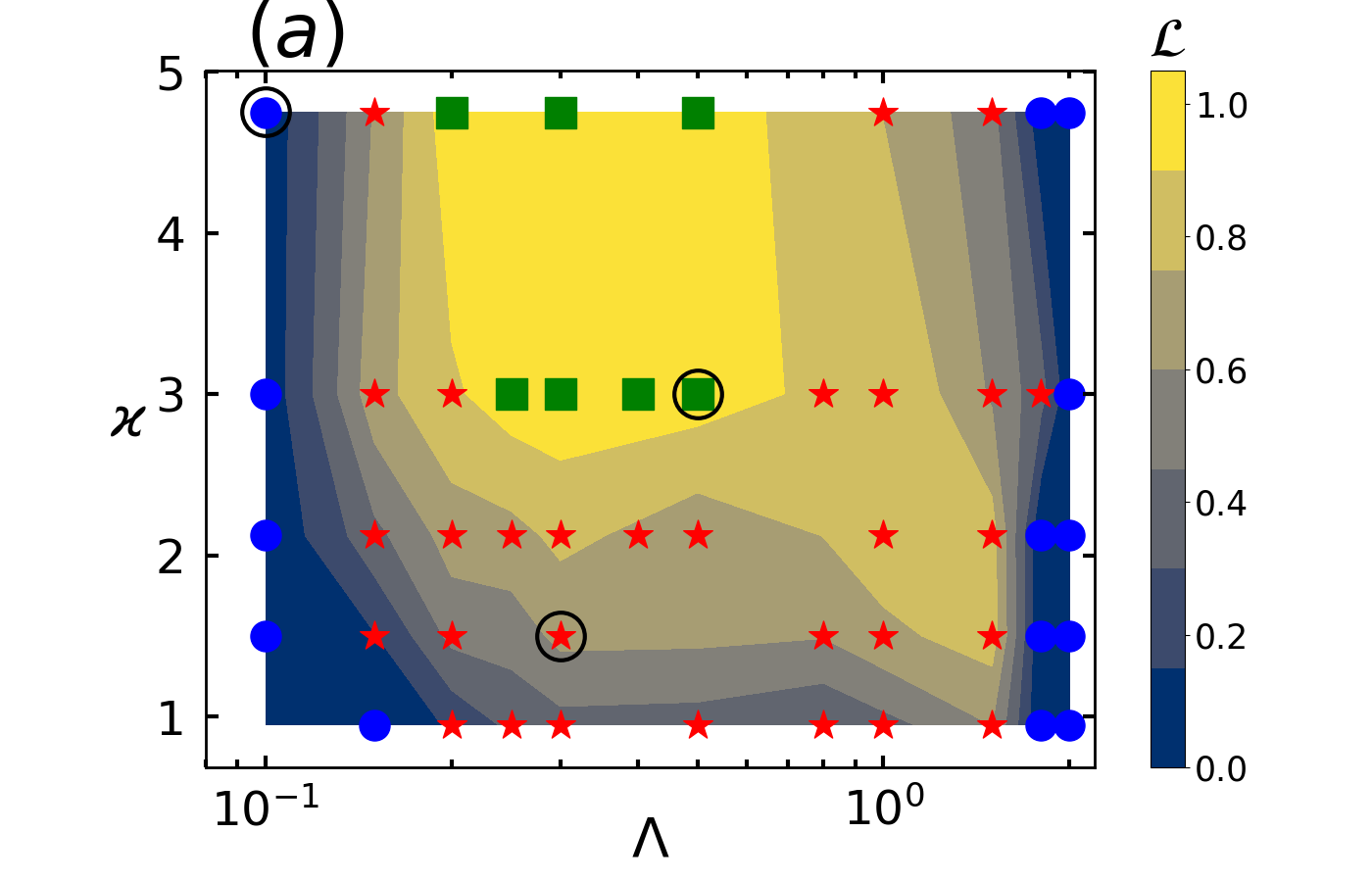}
  \includegraphics[width = 0.2\linewidth, height=0.25\linewidth]{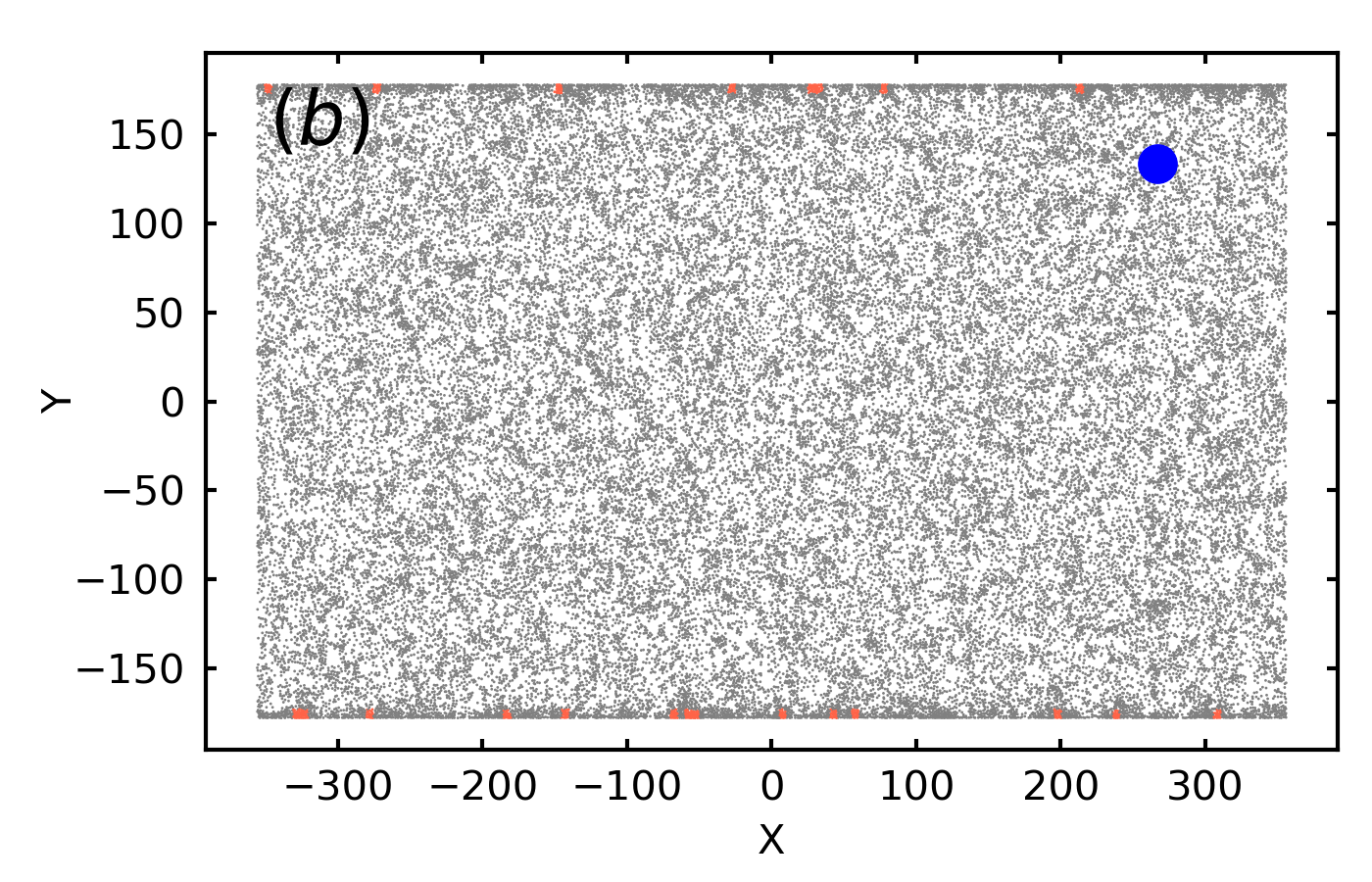}
  \includegraphics[width = 0.2\linewidth, height=0.25\linewidth]{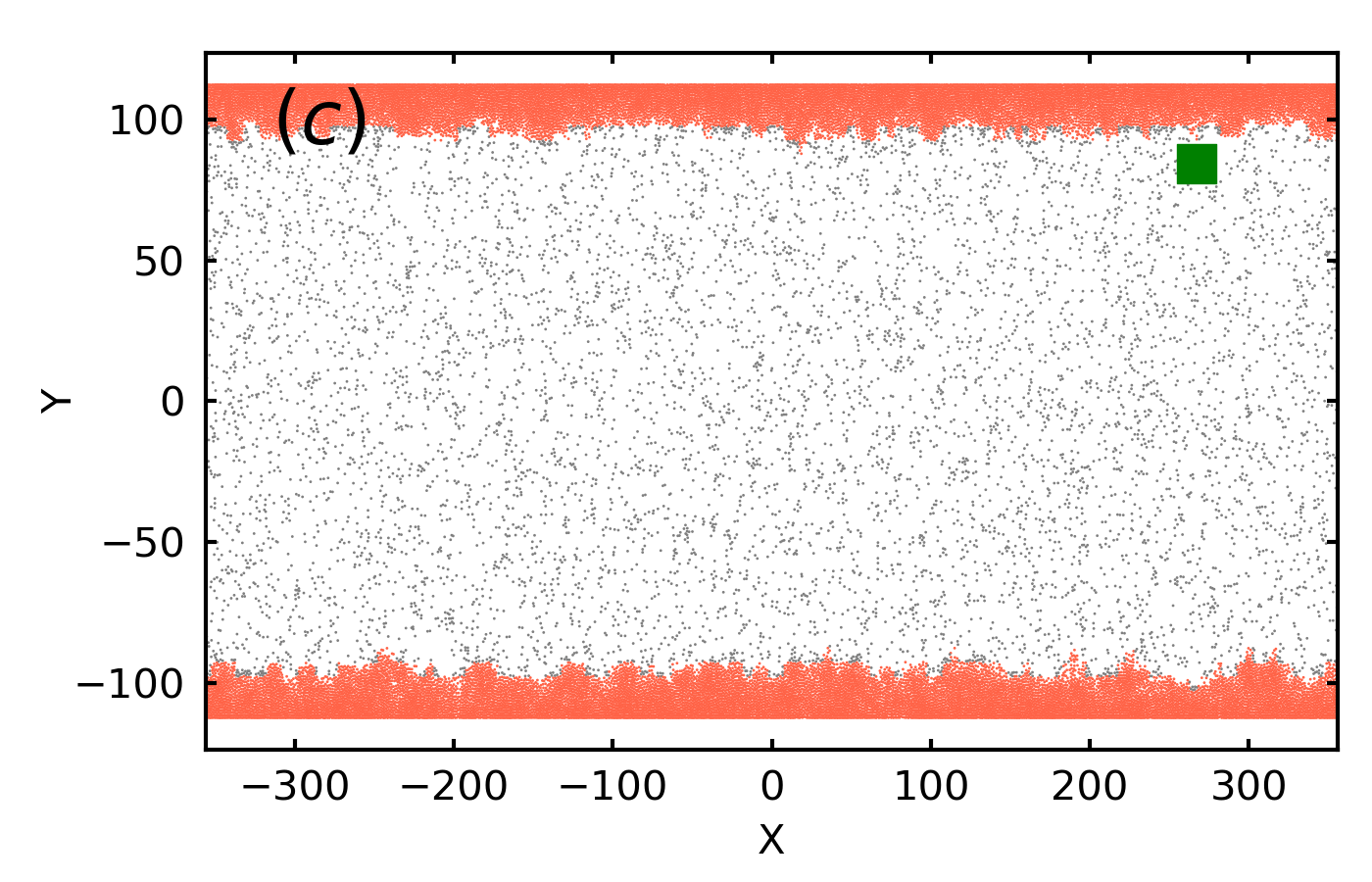}
  \includegraphics[width = 0.2\linewidth, height=0.25\linewidth]{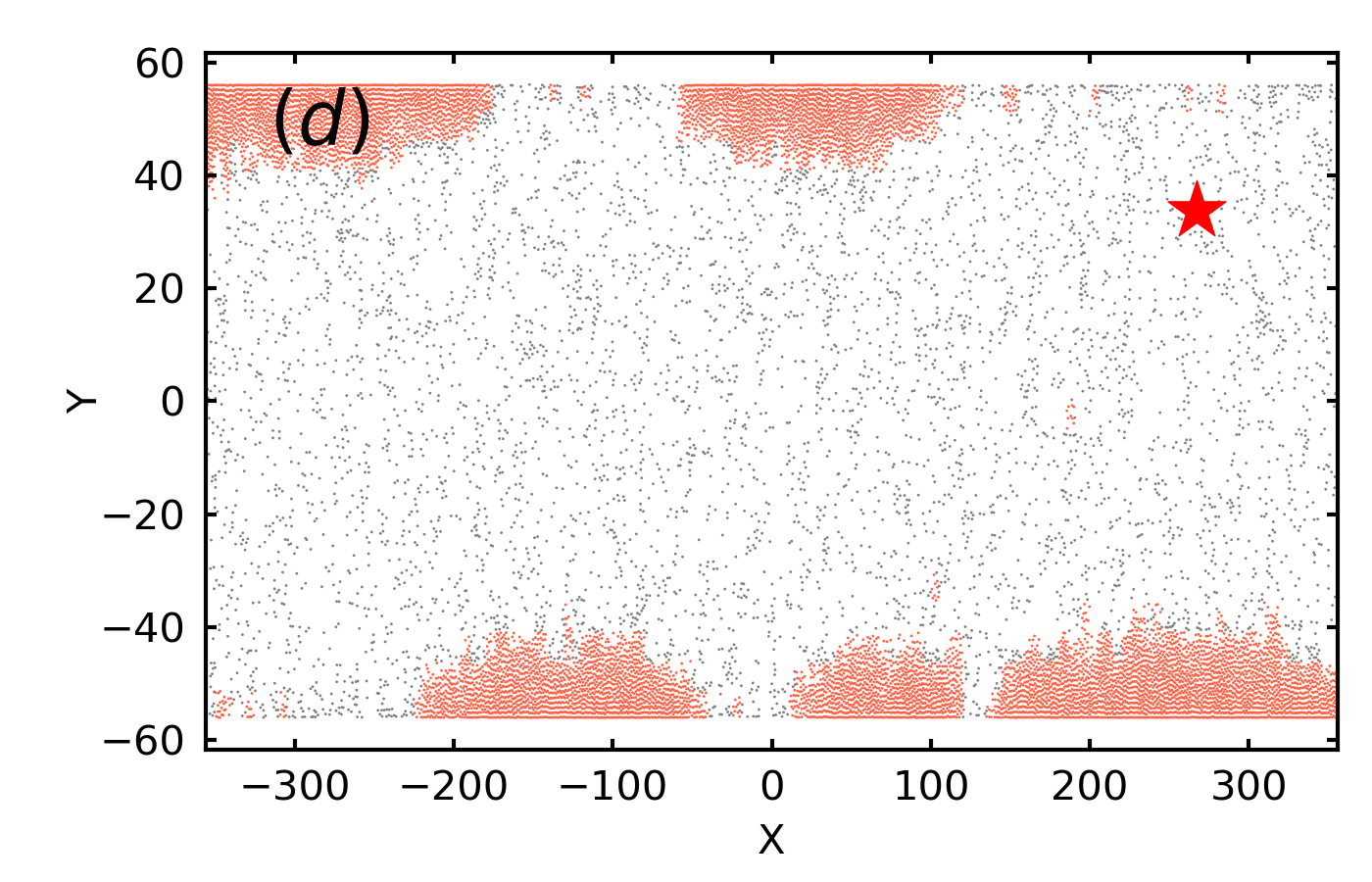}
  \caption{\label{fig:fig2}(a) Phase Diagram in the parameter space of channel width, $\varkappa = L_y / \sigma P_e$ and alignment strength, $\Lambda$. Three different states can be distinguished as (b) homogeneous state (blue circle) with $\mathcal{L} \leq 0.1$, (c) uniform aggregation state (green rectangle) with $\mathcal{L} \geq 0.9$ and (d) non-uniform aggregation state (red star) with $0.1 < \mathcal{L} < 0.9$.}
\end{figure*}
We study a system of self-aligning soft active disks of effective diameter $\sigma$ in a semi-confined box of aspect ratio $L_x / L_y$ with periodic and open boundary conditions along $x$ and $y$-directions respectively. The disks are confined along $y$-direction using potential form \cite{rana2019tuning} $W(y) = W_0 (1 + tanh \hspace{0.1cm} q(y - 0.5 L_y))\forall y \geq 0$ and $W(y) = W_0 (1 - tanh \hspace{0.1cm} q(y + 0.5 L_y))\forall y < 0$ (Fig. \ref{fig:fig1}) with the strength $W_0$ such that it is greater than the individual particle's total energy, ensuring confinement between $\pm 0.5 L_y$ boundary. The parameter $q$ controls the steepness of the boundary. $q$ is fixed to a large value ($q = 10$) to mimic the hard wall limit where only the closest layer to the boundary experiences large repulsive force. 

The self-alignment mechanism is same as described in Ref\cite{de2022reentrant} where each disk of mass $m$ has two sub-hemispheres : head and tail with respective force centers $2l_0$ distance apart. The head and tail of the disks separately interacts with parts of the other disks with a modified Yukawa potential $U^{ht}_{ij} = U^{ht}_0 \displaystyle{e^{-(r^{ht}_{ij}-\sigma)/ \lambda/r^{ht}_{ij}}}$ . $\lambda$ controls the softness of the disks and the criteria, $\tau_0 = l_0(U^{tt}_0 - U^{hh}_0) > 0$, non-reciprocally aligns the disks along the inter-particle separation and towards each other. Fig. \ref{fig:fig1} shows an illustrative diagram of particle-wall interaction and particle-particle interaction. Using an in-house developed Molecular Dynamics code MPMD\cite{de2022collective, de2022motility, de2022reentrant}, we solve the following set of underdamped Langevin equations.  
\begin{align}
  m \ddot{\textbf{r}_i} & = -\gamma \dot{\textbf{r}_i} + \sqrt{2 \gamma^2 D} \pmb{\xi}_i + \sum_j \textbf{F}_{ij} + \gamma \textrm{v}_0 \textbf{n}_i - \pmb{\nabla}W(y_i), \label{eqn:eqn1}\\
   \dot{\textbf{n}}_i & = \sqrt{2 D_r} \pmb{\zeta}_i \times \textbf{n}_i- + \frac{1}{\gamma_r} \sum_j \textbf{T}_{ij} \times \textbf{n}_i. \label{eqn:eqn2}
\end{align}
where $\textbf{F}_{ij} = U_0 \frac{e^{-(r_{ij}-\sigma)/\lambda}}{r_{ij}^2} (\frac{1}{r_{ij}} + \frac{1}{\lambda}) \textbf{r}_{ij}$ is the resultant force experienced by the center of mass of the particle $i$ due to its head and tail component's soft interaction with the head and the tail component of the particle $j$ in limiting approximation, $r_{ij} \gg 2l_0$. $\textbf{T}_{ij} = \frac{\tau_0}{U_0} \textbf{F}_{ij} \times \textbf{n}_j$ is the torque that naturally appears out of the resultant soft interactions on the force centers head and tail component of the particles and $U_0$ is the resultant interaction strength given by $U_0 = U^{hh}_0 + 2 U^{ht}_0 + U^{tt}_0$. The derivation of the resultant force, $\textbf{F}_{ij}$ and the torque $\textbf{T}_{ij}$ can be found in Ref \cite{de2022reentrant}. $\textbf{n}_i$ denotes the propulsion direction pointing towards the head and perpendicular to the line dividing the disk into two halves. Here, $\gamma$ and $\gamma_r$ are dissipation coefficients, $D$ and $D_r$ are diffusion coefficients. The suffix r denotes rotational parameter. $\pmb{\xi}_i$ and $\pmb{\zeta}_i$ are Gaussian white noise. We measure length, time and energy in units of $\sigma$, $1/D_r$ and $k_BT$ respectively. Therfore, the reduced parameters are the inertial parameter, $M = \frac{m / \gamma}{1 / D_r}$, interaction strength. $\Gamma = \frac{U_0 / \sigma}{k_BT}$, softness parameter, $\kappa = \sigma / \lambda$, Peclet number, $P_e = \frac{1/D_r}{\sigma / \textrm{v}_0}$ and reorientational interaction strength, $\Lambda = \frac{1 / D_r}{\gamma_r \sigma^2 / \tau_0 }$. The corresponding reduced equations are:
\begin{align}
  M \ddot{\textbf{r}_i} & = - \dot{\textbf{r}_i} + \sqrt{2} \pmb{\xi}_i + \Gamma \sum_j \frac{e^{-\kappa(r_{ij} -1)}}{r_{ij}^2}\left(\frac{1}{r_{ij}} + \kappa \right) \textbf{r}_{ij} + P_e \textbf{n}_i - \pmb{\nabla}W(y_i), \label{eqn:eqn3}\\
   \dot{\textbf{n}}_i & = \left [\sqrt{2} \pmb{\zeta}_i + \Lambda \sum_j \frac{e^{-\kappa(r_{ij} -1)}}{r_{ij}^2}\left(\frac{1}{r_{ij}} + \kappa \right) \textbf{r}_{ij} \right] \times \textbf{n}_i . \label{eqn:eqn4}
\end{align}

A full scale parameter study for $M$, $\Gamma$, $\kappa$, and $P_e$ is reported in \cite{thesisSoumen}. The study reveals that increasing inertia, $M$ and softness (decreasing $\kappa$) suppresses MIPS, while $P_e$ promotes MIPS, enabling its occurrence even at relatively low values of $\kappa$. Additionally, as $\Gamma$ increases, MIPS exhibits re-entrant behavior for a given $\kappa$, $M$ and $P_e$. In the present study, we consider soft disks with finite inertia and repulsive interactions of intermediate strength. Therefore, the considered fixed parameters are: $M = 0.05$, $\Gamma = 25.0$, $\kappa = 5.0$, $P_e = 75.0$. We define $\varkappa = L_y / l_p$ as the channel width and area fraction, $\phi = N \pi \sigma^2 / (4 Lx L_y)$ is fixed to $0.15$ throughout the study. $l_p$ is the persistence length of the active particle which in terms of Peclet number can be written as $l_p = \sigma P_e$. Thus, $\varkappa = L_y / (\sigma P_e)$. In the absence of the alignment strength, system remains homogeneous at the chosen value of the packing fraction. The system variables are alignment strength, $\Lambda$ and channel width, $\varkappa$ which signifies the strength of orientation of the self-propulsion disks towards each other and the narrowness of the pipe geometry respectively. For the given fixed parameters, a 2D doubly periodic system displays re-entrant phase-separation behavior, as the system reenters into homogeneous phase via MIPS phase upon increase in alignment strength\cite{de2022reentrant}. We study the dynamics of $N = 48400$ disks in the parameter space $\varkappa - \Lambda$ with confinement along $y-$direction resulting in a 2D box system same as rectangular pipe geometry. All the simulations are performed for 5000 units of time with step size, $\delta t = 0.0005$ and the averaged quantities are summed over 200 steady states.

\section{\label{sec:Results}Results and discussion}
\begin{figure}[h]
  \includegraphics[width = 0.8\linewidth, height=0.55\linewidth]{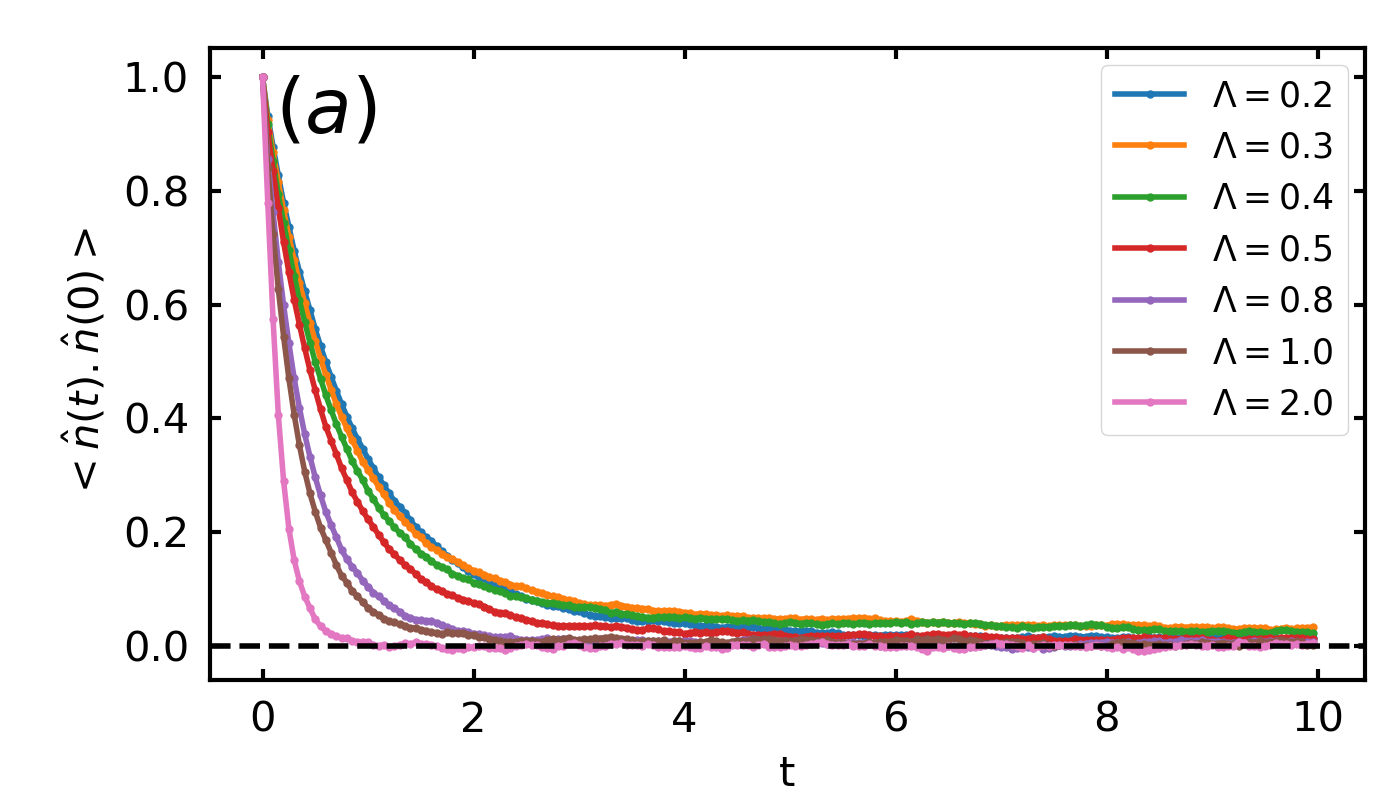}
  \includegraphics[width = 0.48\linewidth, height=0.4\linewidth]{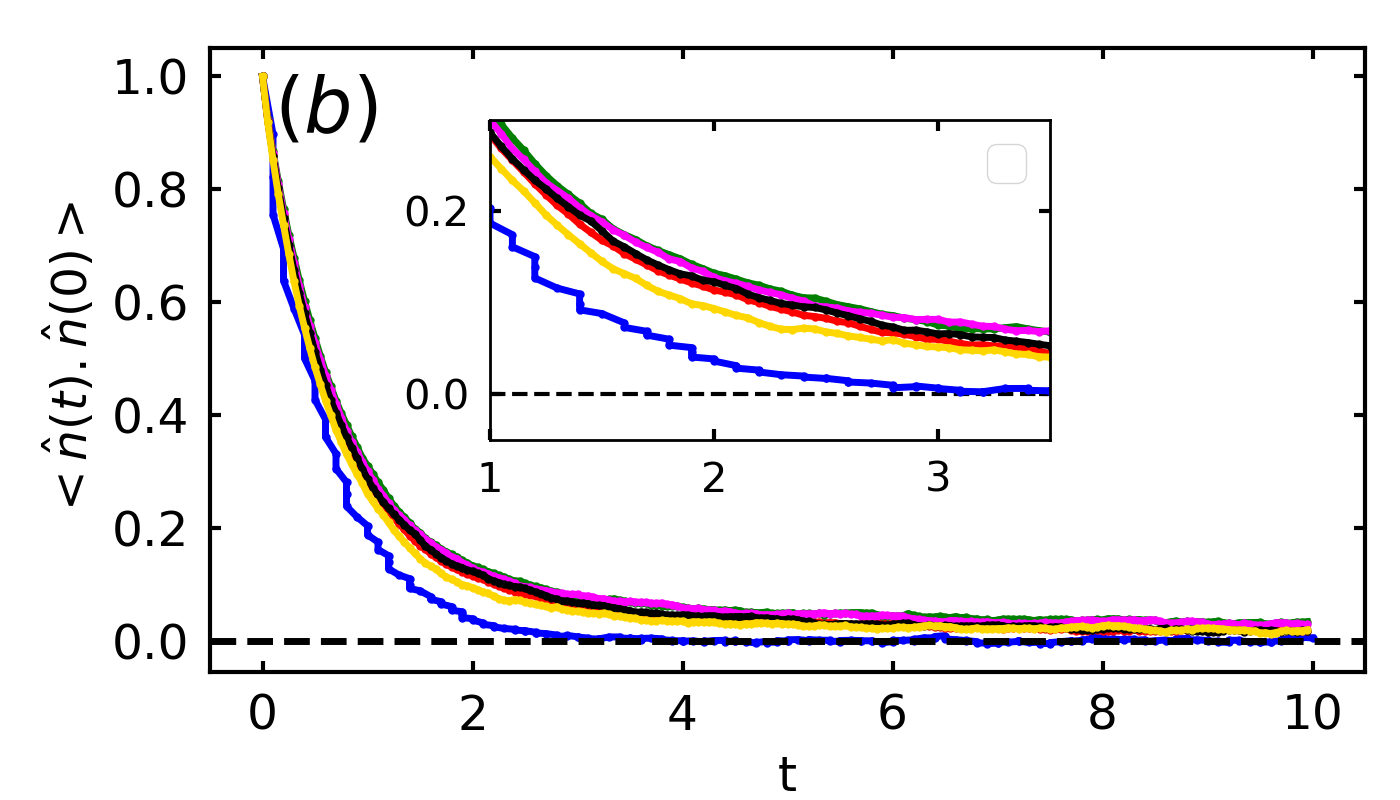}
  \includegraphics[width = 0.48\linewidth, height=0.4\linewidth]{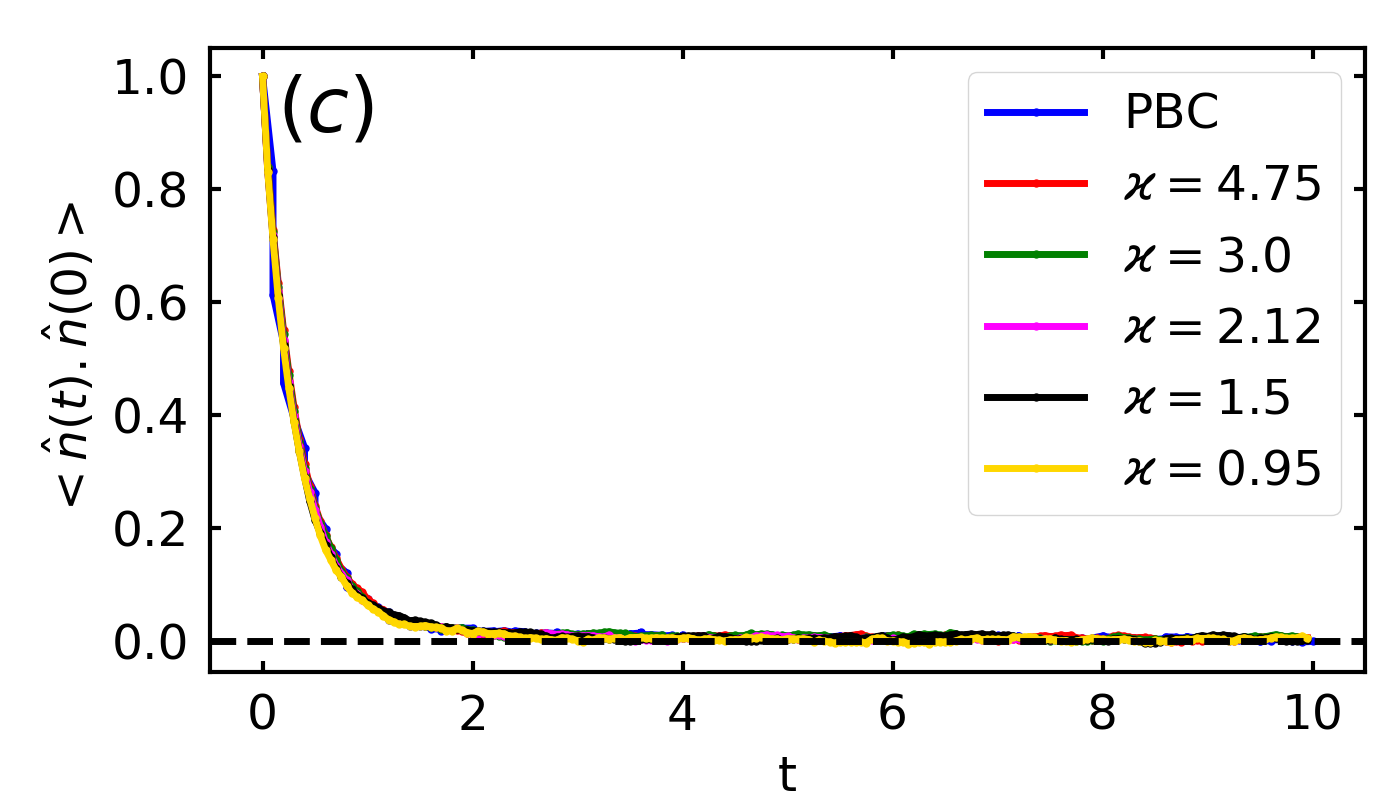}
  \caption{\label{fig:fig3} (top) Autocorrelation of the self-propulsion direction $<\mathbf{n}_i(t) . \mathbf{n}_i (t = 0)>$ for various values of $\Lambda$ at fixed $\varkappa = 3.0$. (bottom) Comparison of the Autocorrelation of the self-propulsion direction in semi-confined geometry at various $\varkappa$ with doubly periodic boundary system (blue line) at fixed (left) $\Lambda = 0.3$ and (right) $\Lambda = 1.0$.} 
\end{figure}
\begin{figure}[h]
  \includegraphics[width = 0.99\linewidth, height=1.1\linewidth]{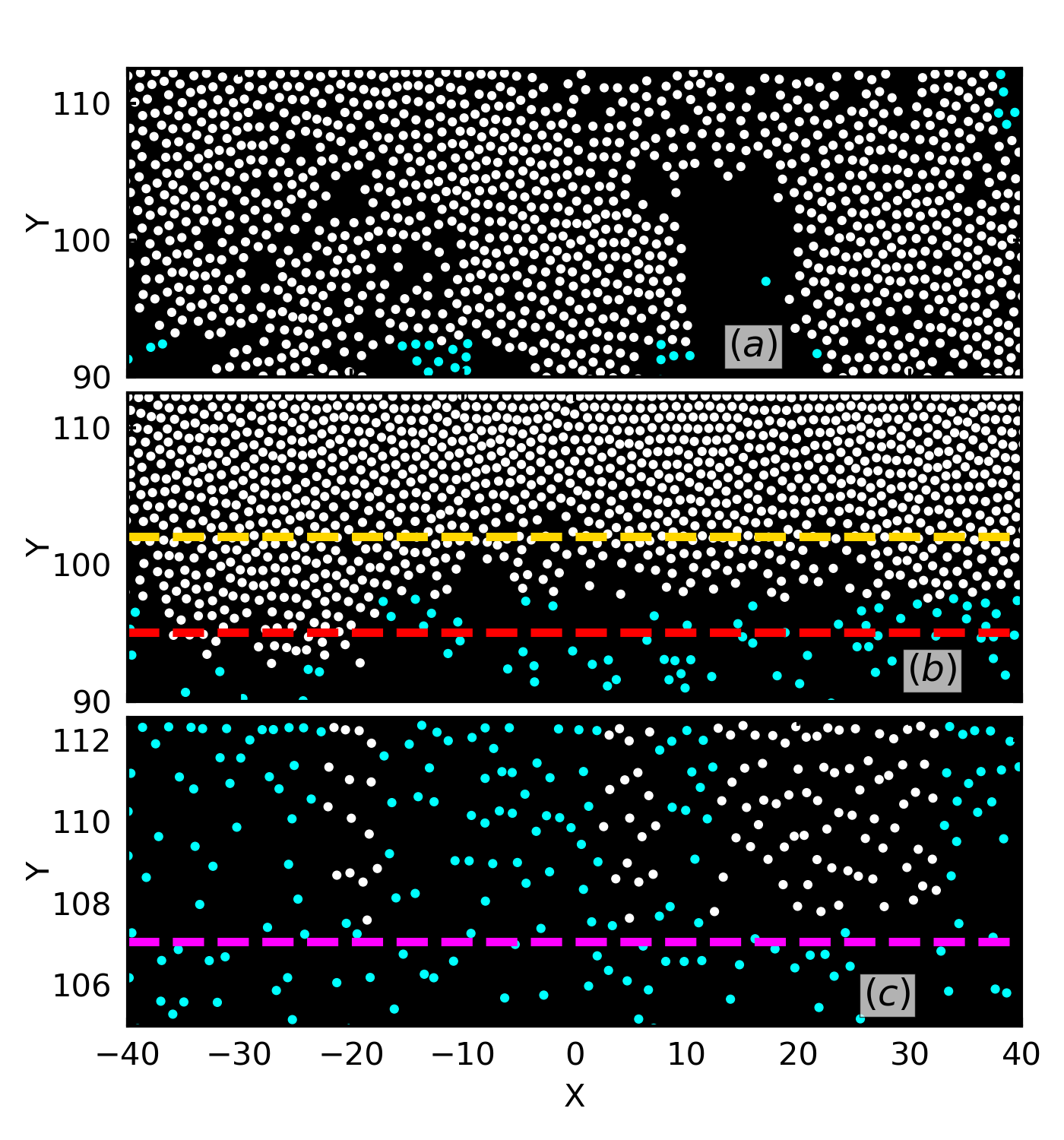}
  \includegraphics[width = 0.99\linewidth, height=0.45\linewidth]{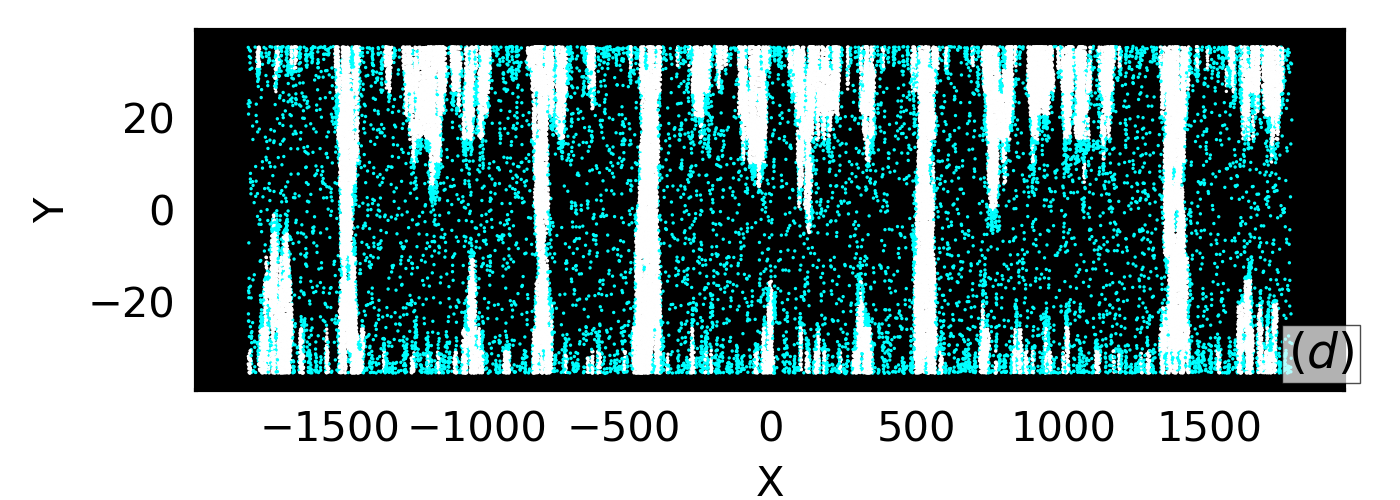}
  \caption{\label{fig:fig4} Configuration of particles for (a) $\Lambda = 1.0$, (b) $\Lambda = 0.3$ and (c) $\Lambda = 0.1$ at fixed $\varkappa = 3.0$. (d) Particles' configuration for $\varkappa = 0.95$ and $\Lambda = 1.0$. Particles in dense and dilute phase are shown in white and cyan respectively. Layered and non-layered structures are observed for intermediate and high $\Lambda$ values respectively. Dashed lines corresponds to a particular distance from the wall as shown in Fig.\ref{fig:fig5}c. Magenta and red dashed lines are positions of average height of the structure and yellow dashed line is the minimum height of the structure. Only a fraction of geometry is shown for configurations in (a), (b) and (c).} 
\end{figure}
\begin{figure}[h]
  \includegraphics[width = 0.48\linewidth, height=0.4\linewidth]{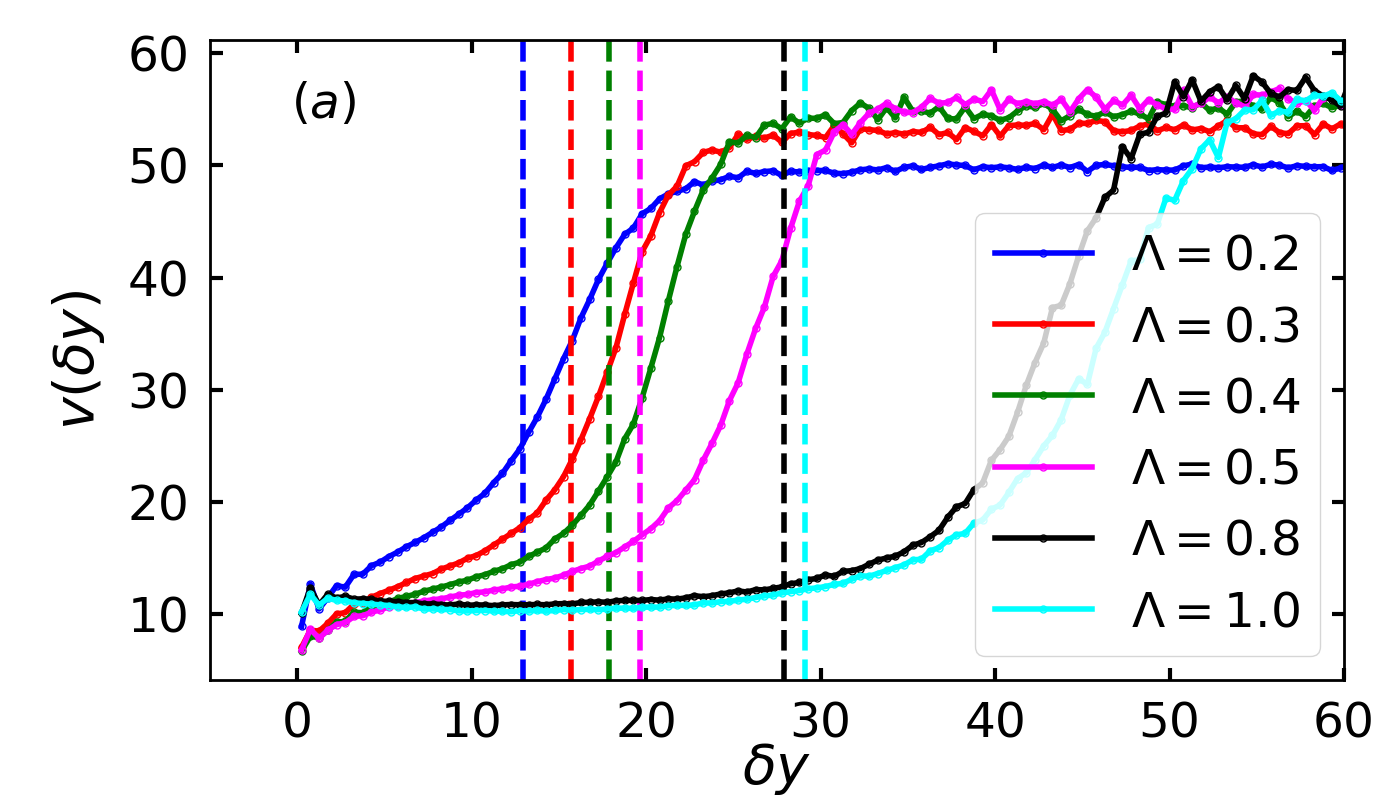}
  \includegraphics[width = 0.48\linewidth, height=0.4\linewidth]{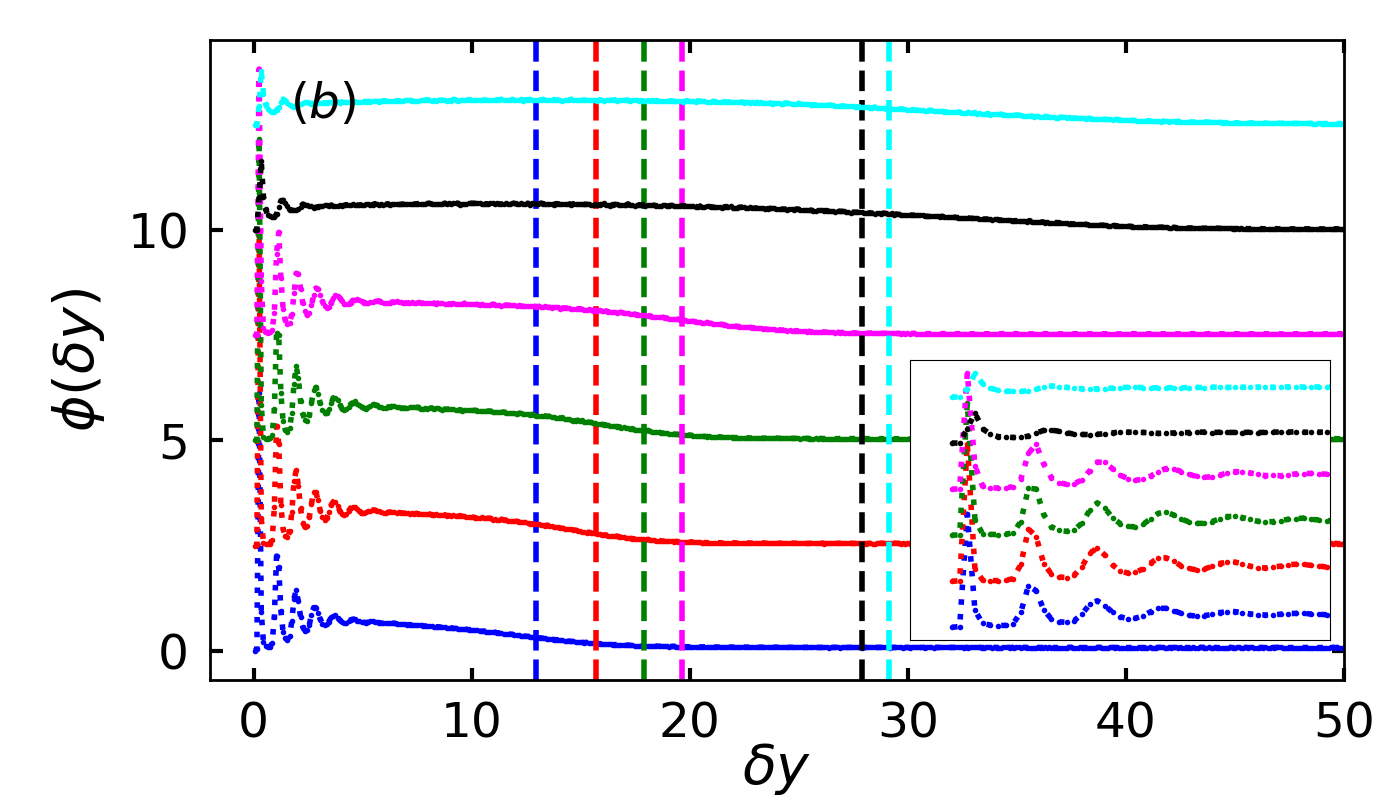}
  \includegraphics[width = 0.99\linewidth, height=0.7\linewidth]{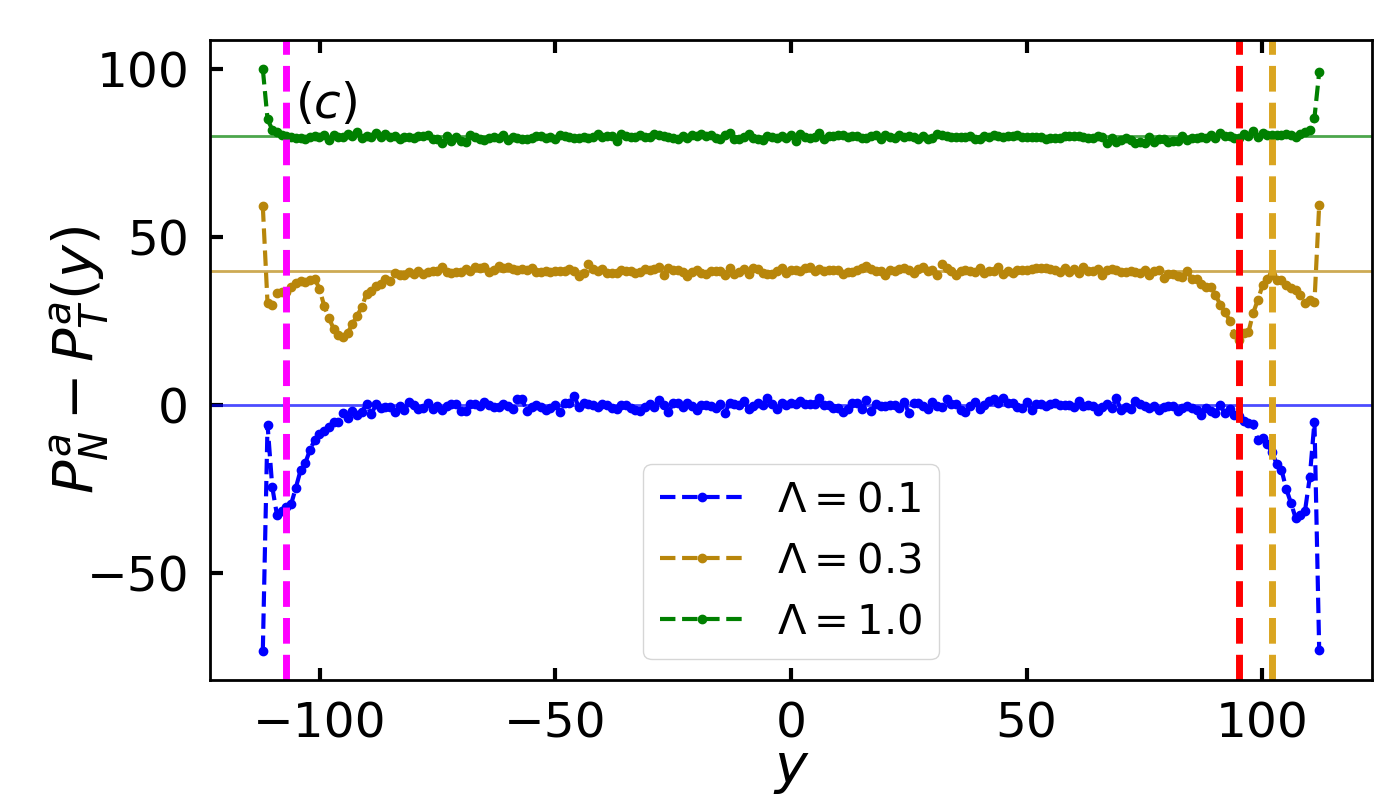}
  \caption{\label{fig:fig5}(a) Local speed, $\textrm{v}$ and (b) local area fraction, $\phi$ as a function of particles' distance from the wall boundary, $\delta y$ for fixed channel width, $\varkappa = 3.0$. Inset shows a zoomed portion of the main plot to highlight its peaks. (c) Active Pressure difference between the normal and the tangential components, $P^a_N - P^a_T$ as a function of distance from the wall, $\delta y$ for different values of alignment strength, $\Lambda$ and fixed channel width, $\varkappa = 3.0$. Pressure curves are shifted along $y$ by 40 units for representative purpose. } 
\end{figure}
\begin{figure}[h]
  \includegraphics[width = 0.9\linewidth, height=0.48\linewidth]{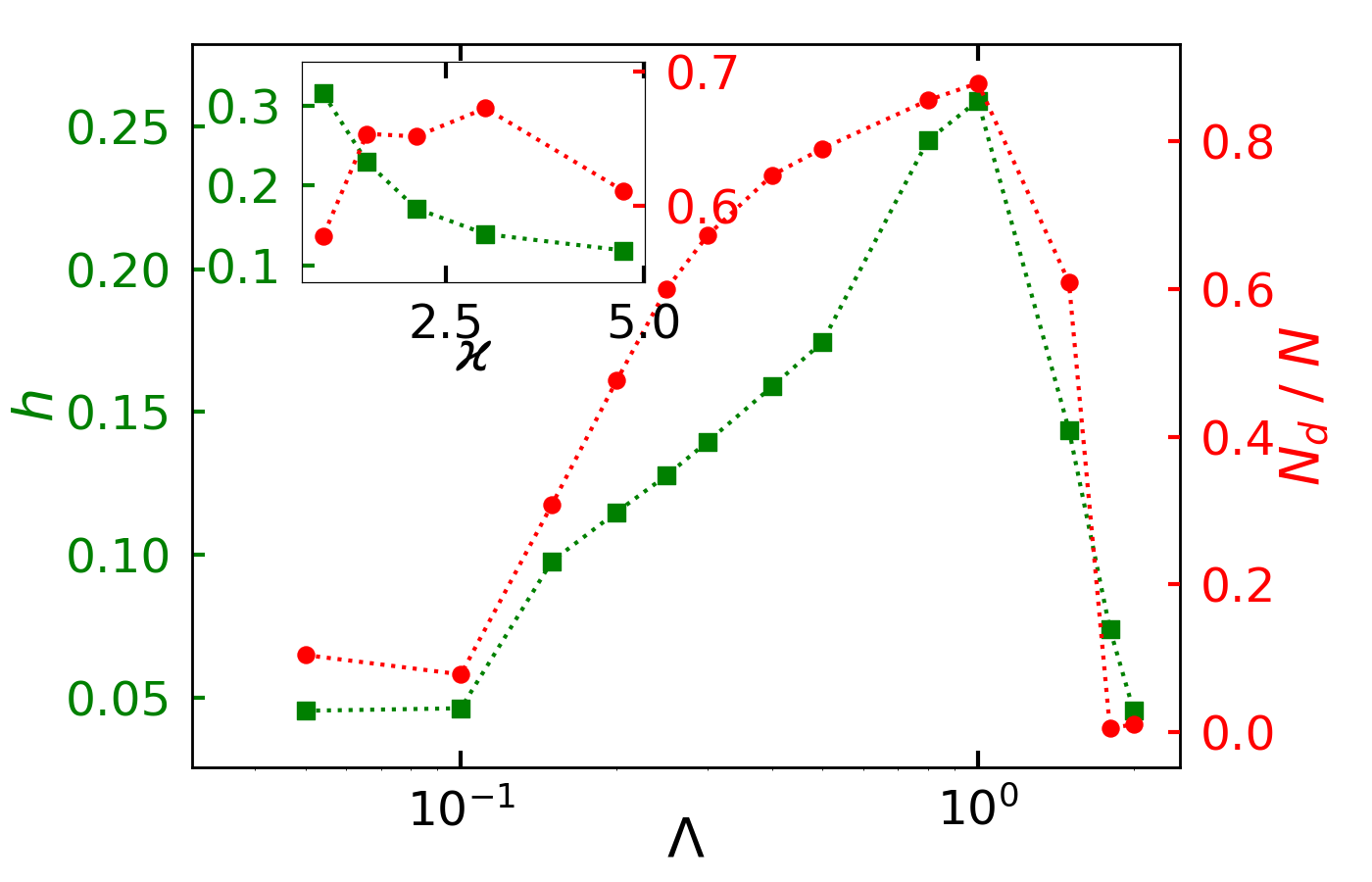}
  \caption{\label{fig:fig6} Wall aggregate height, $h$ (green circle) and cluster fraction, $N_d / N$ (red circle) as a function of alignment strength, $\Lambda$ for fixed channel width, $\varkappa = 3.0$. Inset shows the variation of these quantities with channel width, $\varkappa$, for a given alignment strength, $\Lambda = 0.3$.} 
\end{figure}

Fig.\ref{fig:fig2}a summarises the effect of the channel width, $\varkappa$ and alignment strength, $\Lambda$ in the form of a phase diagram between $\varkappa$ and $\Lambda$. This phase diagram is divided into distinct regions according to the value given by aggregation length, $\mathcal{L}$. Aggregation length, $\mathcal{L}$ serves as an identifier for different aggregates morphology (see Appendix \ref{AppendixA} for the definition and calculation). For negligible accumulation of particles along the walls, where aggregation length , $\mathcal{L} \leq 0.1$, the state is called homogeneous state (Fig. \ref{fig:fig2}b). When the particles form clusters along the whole length of the wall, the state is called uniform aggregation state where aggregation length , $\mathcal{L} \geq 0.9$ (Fig. \ref{fig:fig2}c) and when the accumulation or the formation of clusters is at specific sites on the wall leaving a portion of the wall unoccupied of clusters, then this state is called non-uniform aggregation state and $\mathcal{L}$ lies between $0.1$ and $0.9$ (Fig. \ref{fig:fig2}d). We observe that the system displays re-entrant behaviour where the system is found in homogeneous state at low and high alignment strength which is similar to what is observed with doubly periodic boundaries\cite{de2022reentrant}. However for confined systems, cluster formation occurs at wall boundaries as the wall acts as a potential nucleation site for the formation of the cluster. Particles approaching the wall get stuck there until they turn their self-propulsion direction away from the wall. This effectively slows down the particle and ensues wall accumulation. For low $\Lambda$, the nucleation occurs only at the wall boundaries. As the alignment strength increases, large number of small clusters appear in between the walls which eventually dissolves and gets accumulated at the wall boundaries. Fig. \ref{fig:fig2}a shows that at low and high values of the alignment strength, the occurrence of homogeneous state is independent of the channel width, $\varkappa$. Our results demonstrate that the formation of a cluster or aggregate is linked to the alignment between the disks while location or the spread of the cluster is dominated by the interplay of alignment strength and the channel width.

Uniform aggregation state is observed at intermediate values of $\Lambda$ and for $\varkappa > 2.0$. With decrease in channel width, the pipe geometry becomes narrow which increases the effective collisions between the disks along the path perpendicular to the wall. Hence, due to this space in-homogeneity of the collisions where the probability of collisions perpendicular to the wall is greater than the collisions in the direction parallel to the wall and the morphology of the aggregates tend to change to non-uniform aggregation state. This effect can be seen in Movie1 and 2 where the shape of the transient clusters shifts from circular to elongated with decrease in channel width and increase in alignment strength as the particles joins from above and below rather than particles joining the cluster uniformly from all sides. For the given $P_e$, with decrease in channel width, particles reach the wall faster which increases the probability of collisions with the particles on the way towards the wall. Hence, the nucleation sites grow quickly to exhibit in-homogeneous particle accumulation at the walls. Whereas the particles get more time due to comparatively lower collision frequencies at larger $\varkappa$ to move out of the transient clusters at the walls and asymptotically form uniform accumulation.

Fig. \ref{fig:fig3} captures the effect of increased collisions upon increase in $\Lambda$ and decrease in $\varkappa$ using the autocorrelation of self-propulsion direction which decays faster upon increase in $\Lambda$ (Fig.\ref{fig:fig3}a) and decrease in $\varkappa$ (Fig. \ref{fig:fig3}b). However, at a sufficiently large $\Lambda$, the strength of alignment torque becomes so large that the decay of the autocorrelation becomes independent of the gap between the walls (Fig. \ref{fig:fig3}c). Furthermore, we also compare the autocorrelation in this confined geometry at various channel widths to the autocorrelation in doubly periodic system (PBC) for small and large values of $\Lambda$ (Fig.\ref{fig:fig3}b, \ref{fig:fig3}c). It is observed that the autocorrelation decays slower in comparison to doubly periodic system (PBC) for small $\Lambda$ and the decay rate approaches to that of doubly periodic system (PBC) at large $\Lambda$ as shown in Fig. \ref{fig:fig3}b and Fig. \ref{fig:fig3}c respectively. This can be explained on the basis of layered particle arrangement as discussed in the paragraphs to follow.

Fig. \ref{fig:fig4} shows distinct structural patterns at different values of $\Lambda$ for a given $\varkappa = 3.0$ and  for $\Lambda = 1.0$, $\varkappa = 0.95$. Depending on the local arrangement of particles, these structures are distinguished as layered and non-layered structures. At intermediate values of $\Lambda$, we observe layered structure (Fig. \ref{fig:fig4}b) while at larger $\Lambda$, the layering vanishes (Fig. \ref{fig:fig4}a). Due to increased alignment strength between the particles, a large number of small dynamic clusters appear in between the walls and these clusters tend to join the clusters at the wall via a bridge of clusters closest to them which leads to random local arrangement of the particles also leaving gaps or dilute regions in between dense regions as shown in Fig. \ref{fig:fig4}a. Thus, layering of particles close to the wall does not happen. For $\varkappa < 1$, these elongated clusters from the upper and the lower wall tend to join and form lane like structures as shown in Fig. \ref{fig:fig4}d. At low alignment strength, wall accumulation of the particles dominate over cluster formation in between the walls. Particles approaching the wall get trapped until they change their self-propulsion direction away from the wall, while the particles leaving the wall can move freely. This asymmetry leads to wall accumulation and slower decay of self-propulsion direction autocorrelation in comparison to doubly periodic systems as shown in Fig. \ref{fig:fig3}b. Thus, particles at low alignment strength tend to form layers.

To visualise the the formation of layers close to the wall, we plot local packing variation as a distance from the wall as shown in Fig. \ref{fig:fig5}b. The peaks in the curve are reflective of the layered structure, which is present at small $\Lambda$ and vanishes at large $\Lambda$, when the alignment strength or torque takes over the wall slow down of the particles. The colored dashed line shows the position of the average height of the clusters of respective $\Lambda$. In layered structure, particles are trapped and not free to move, while for non-layered structures the particles are highly mobile which leads to the thermalisation of particles. Thus, for layered structure, i.e. at low $\Lambda$, the variation of speed as a distance from the wall is gradual and increases towards dilute-dense interface or the average height of the structure, while for non-layered structure, where particles can move freely, the speed variation is nearly constant upto a distance of average height of structure and increases in the dilute region. In the succeeding paragraphs, we show that active pressure difference is a useful
indicator for different aggregate morphologies and the peaks in the pressure curve provide valuable information about the average and the minimum height of the structure.

Fig. \ref{fig:fig5}c shows pressure difference curve for different values of $\Lambda$ for a given $\varkappa = 3.0$. The contribution of the pressure is due to the motility of active particles and is given by $P^a_{\alpha \beta} (y) = \frac{1}{2 L_x \delta y} < \sum_{i \in \delta y} j_{i\alpha} \textrm{v}_{i\beta} > $ where $< . >$ is time averaged within rectangular bins parallel to the wall and of width $\delta y = 1$. $\textbf{j}_i = \gamma \textrm{v}_0 \textbf{n}_i / D_r$ stands for active impulse produced by active force $\gamma \textrm{v}_0 \textbf{n}_i$ and $\textrm{v}_i$ is the velocity of the $i$th particle. The non-zero diagonal components, $P^a_{xx}$ and $P^a_{yy}$ are respectively the tangential ($P^a_T$) and normal ($P^a_N$) components of the pressure tensor, $P^a_{\alpha \beta}$. However, calculating the difference between the normal and tangential components of pressure is of practical interest as one can obtain surface tension by evaluating the integrated difference of these components \cite{kirkwood1949statistical}. Active pressure difference, $P^a_N - P^a_T$ curve gives us the information about the layered structure, where the dips in the pressure curve are indicative of layered arrangement of particles (yellow curve in Fig. \ref{fig:fig5}c). The longer dip (red dashed line) in the pressure difference curve gives the average height of the structures or the average distance of dense dilute interface and smaller dip (yellow dashed line) provides the minimum height of the structures. The absence of peaks in the pressure difference curve (green curve in Fig. \ref{fig:fig5}c) suggests the absence of layers in the structures. Furthermore, for layered structures, depending on whether the pressure difference for the layer closest to the wall is positive or negative, we observe uniform aggregation and non-uniform aggregation states respectively.

\begin{figure}[h]
  \includegraphics[width = 0.98\linewidth, height=0.9\linewidth]{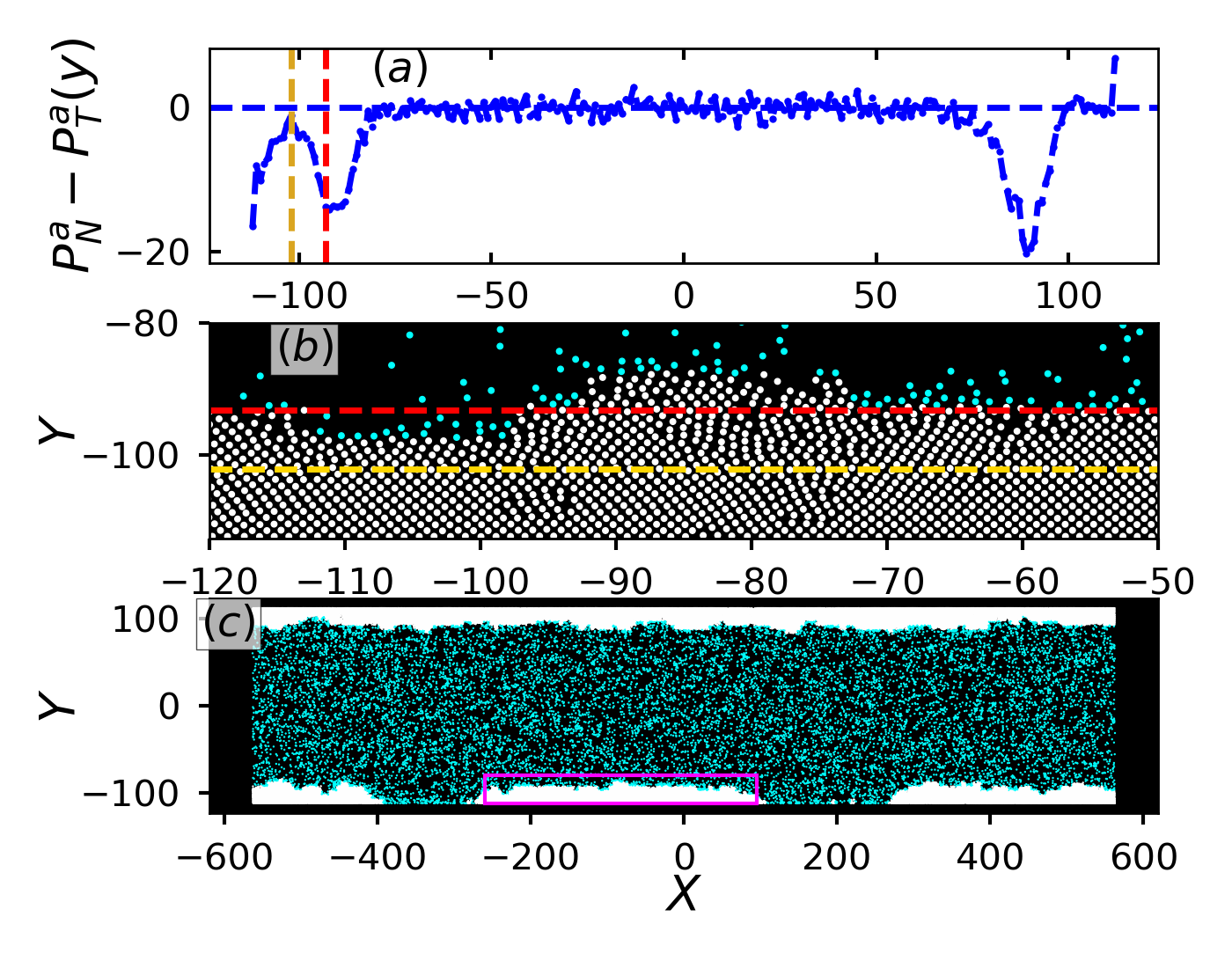}
  \includegraphics[width = 0.98\linewidth, height=0.55\linewidth]{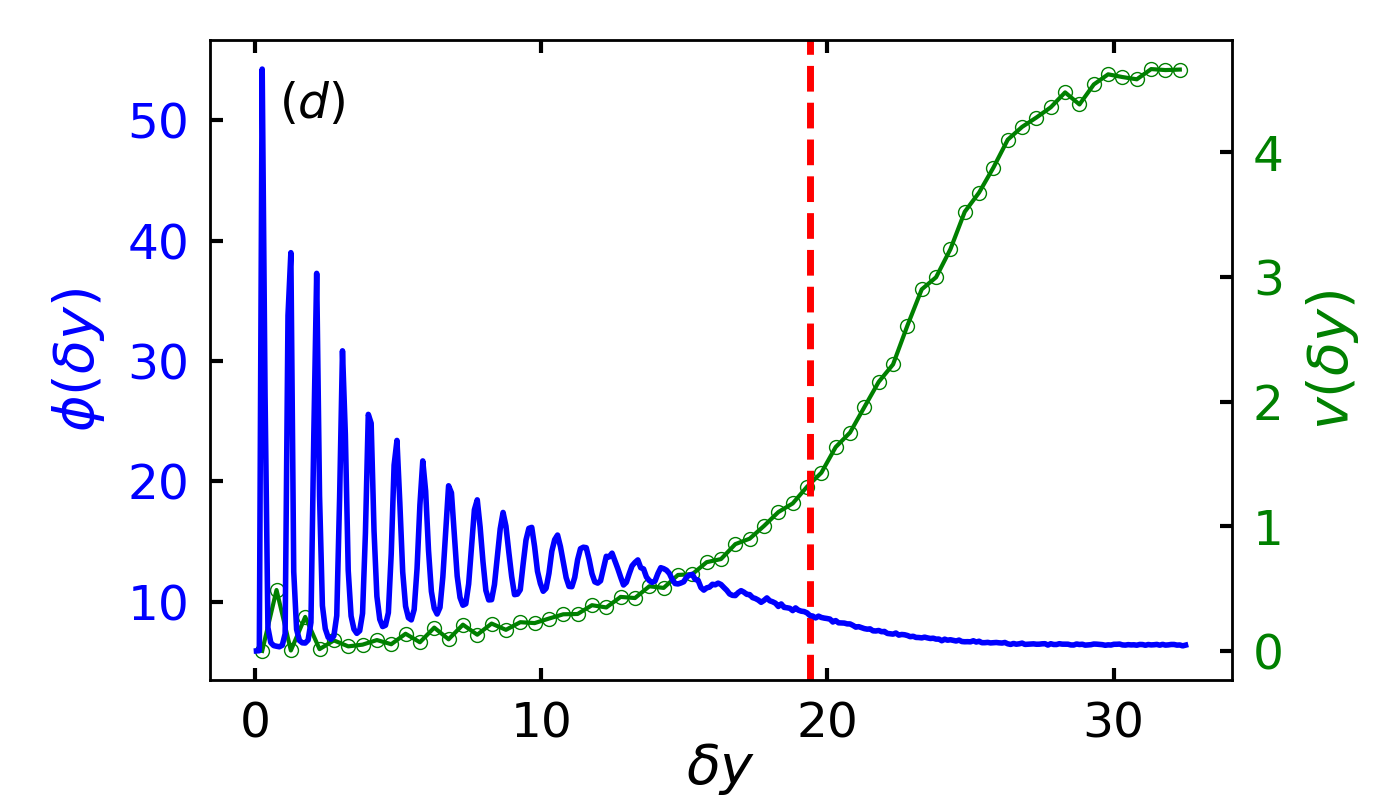}
  \caption{\label{fig:fig7} (a) Active Pressure difference, $P^a_N - P^a_T$ as a function of distance from the wall, $\delta y$ for $\Lambda = 0.2$, $\varkappa = 3.0$ and $\kappa = 18.0$ (hard particle limit). (b) Corresponding particle configuration at the lower wall of the box geometry. (c) local speed, $\textrm{v}$ and local area fraction, $\phi$ as a function of distance from the wall. Peaks in $\phi(\delta y)$ are indicative of layered structure.} 
\end{figure}

Fig. \ref{fig:fig6} shows the steady-state properties such as average height of the wall aggregation, $h$ (See Appendix \ref{AppendixA} for details) and the fraction of particles in the dense cluster, which is denoted by $N_d / N$ and named as cluster fraction. As with increase in $\Lambda$, the number of transient clusters increase which eventually merge into large cluster at the wall boundary, therefore the fraction of particles in dense phase, $N_d/ N$ increases with $\Lambda$. Since the height of wall aggregates is directly linked to the the cluster formation, therefore we observe a simultaneous increase in $h$ and $N_d / N$. With further increase in $\Lambda$, the reorientation frequency becomes faster hence transient clusters form and break rapidly and only a fraction of them are able to reach the wall leading to decrease in height of the wall aggregates. Inset of Fig. \ref{fig:fig6} shows the effect of channel width on the steady state properties for a given alignment strength, $\Lambda = 0.3$. Due to elongated clusters with decrease in channel width, the average height of the wall aggregate increases monotonically with decrease in channel width. On the other hand, the variation of cluster fraction, $N_d / N$ is non-monotonic with decrease in channel width.

Until now, the results pertained to the soft particle limit, whereas the hard particle limit can introduce structural changes in the aggregate as shown in Fig. \ref{fig:fig7}b and Fig. \ref{fig:fig7}d. Fig. \ref{fig:fig7}c shows a snapshot of the configuration of particles in full box geometry, where the particles are colored white and cyan to belong to dense and dilute phase respectively.  Zoomed configuration of the particles in magenta box (Fig. \ref{fig:fig7}c) is shown in Fig. \ref{fig:fig7}b. It is observed that the particles are arranged in a layered fashion. As previously stated, we can quantify the layering effect by plotting the variation of local packing fraction with distance form the wall as shown by the blue curve in Fig. \ref{fig:fig7}d. The $\phi$ curve shows several peaks which is indicative of the layered organisation of particles. Additionally, Fig. \ref{fig:fig7}d exhibits more peaks than Fig. \ref{fig:fig5}b, indicating that softness disrupts the stratification or the layering of the structure. This result is consistent with the fact that the inter-particle distance increases with increase in $\kappa$ or decrease in softness. Moreover, the local speed variation of the particles increases gradually as also observed for other layered structures. Fig. \ref{fig:fig7}a shows the active pressure difference curve where the large dip (red dashed line) is indicative of the average height of the cluster and smaller dip (yellow curve) is the minimum height of the cluster. Positive and negative pressure difference of the layer close to the wall shows uniform and non-uniform accumulation as shown in the full configuration plot in Fig. \ref{fig:fig7}c. This further corroborates the indicative features of the pressure difference curve previously observed for lower $\kappa$.

\section{\label{sec:Conclusions}Conclusions}
In summary, this study investigates the combined effect of particle alignment, $\Lambda$ and confined geometry on the aggregate formation at the walls. We discover different morphologies of particle accumulation which can be categorized as homogeneous state, uniform accumulation state, and noon-uniform accumulation state depending upon the aggregate's length along the wall. The height of the structures display a non-monotonic dependence on $\Lambda$ for a given confinement width, $\varkappa$. The system reenters into homogeneous state upon increase in alignment strength, $\Lambda$ for a given confinement width, $\varkappa$. Interestingly, this re-entrant behavior is independent of the confinement width, $\varkappa$. Different morphologies of the system are explained using auto-correlation of the self-propulsion direction which captures the effect of increased effective collisions between the particles with rising $\Lambda$ and decreasing $\varkappa$. Structural analysis reveals layered formation at low $\Lambda$ and non-layered structures at high $\Lambda$, visualised using local packing fraction variation as a distance form the wall. Moreover, the layered structure information can be extracted using the active pressure difference between the normal and the tangential components. This pressure curve gives insights about the average and minimum height of the structures from the consecutive dips in the curve. In essence, this study advances our understanding of aligned active particle dynamics in confined environments, which may be relevant to biological active particles in channels or natural confined settings.

With a decrease in the steepness parameter $q$, the wall's potential gradient becomes smoother, which can affect layers beyond the layer closest to the wall. Studying the dynamics of the system by adjusting the softness of the wall boundary could be a promising future direction for the present research. Moreover, the geometric configuration of the boundary may be configured in a ratchet-like manner to facilitate the investigation of dynamics within a corrugated channel.


\section*{Acknowledgements}
Numerical simulations in this work has been carried out in ANTYA cluster at IPR.

\bibliography{Manuscript}

\section*{Authors contributions}
S.D.K., R.G. and A.C. designed the simulations. A.C. carried out the simulation runs and generated the data. All the authors contributed in analyzing the data and writing the Manuscript.

\section*{Competing interests}
The authors declare no competing interests.

\appendix
\section{}
\label{AppendixA}
\subsection{Quantification of dense-dilute phase}
\begin{figure}[h]
  \includegraphics[width = 0.45\linewidth, height=0.4\linewidth]{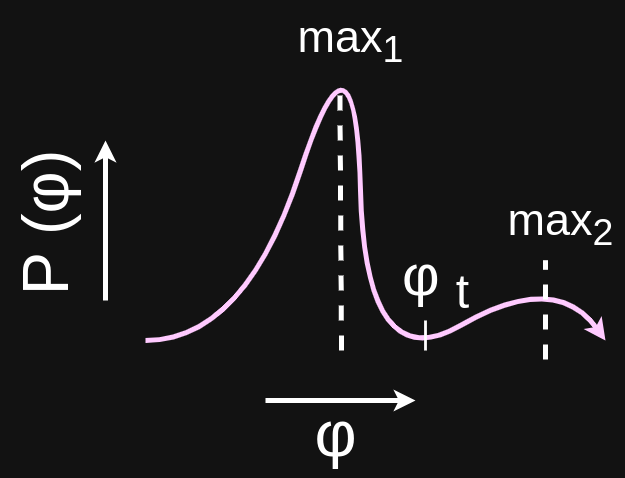}
  \includegraphics[width = 0.45\linewidth, height=0.4\linewidth]{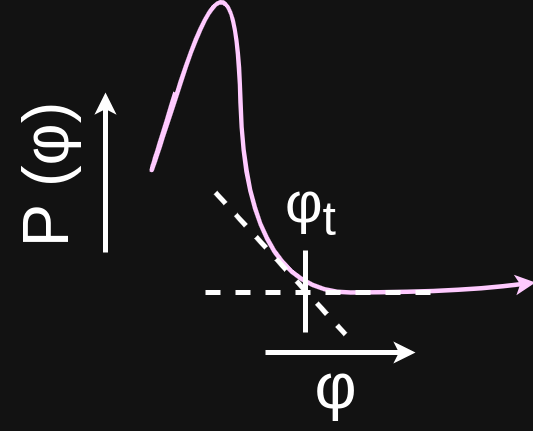}
  \caption{\label{fig:figA1} Criteria for defining target local fraction, $\phi_t$ for (left) double peaked and (right) single peak local packing fraction distribution.} 
\end{figure}

In order to identify the disks that belong to the dense or dilute region, target local packing fraction, $\phi_{t}$ is computed as follows. The simulation box, $L_x \times L_y$ is subdivided into cells of size $5 \times 5$, and the local area fraction of each cell is computed. Using this information and binsize of 1, the distribution of local area fraction is found. To compute $\phi_t$, the following approach is used, depending on whether the distribution curve is single or double peaked. For double-peaked distribution curve, $\phi_t$ corresponds to the point immediately following the first peak's fall and preceding the second peak's rise while for single peak distribution curve, $\phi_t$ is the point close to fall of the peak and the minima value as shown in Fig. \ref{fig:figA1}. The disks belonging to the cell with local packing fraction greater than target packing fraction are tagged as dense particles otherwise they are tagged as dilute particles.

\subsection{cluster fraction ($N_d / N$)}
If the value of the local area fraction of the cell is greater than the target local fraction, $\phi_t$, the disks in that cell belong to the dense phase; otherwise, they belong to the dilute phase. The cluster fraction is now calculated by dividing the total number of disks in the dense phase, $N_d$ to the total number of disks, $N$. We calculate this quantity as a time-averaged quantity over steady states.

\subsection{Aggregation length ($\mathcal{L}$) and height ($h$)}
After tagging the particles as dense and dilute particles corresponding to the dense and dilute regions respectively, we calculate aggregation length, $\mathcal{L}$ and aggregation height, $h$, by dividing the simulation box into cells of size $2 \times 5$. Starting from the top left cell and traversing along cells closest to the wall in x direction, we count the number of cells which are devoid of dense particles. The same process is repeated for the lower wall by starting from the bottom left cell. Computing the fraction of empty cells to the total cells along x and subtracting that from 1 gives the aggregation length. 

To compute aggregation height, $h$, we find the maximum local height of the aggregate or cluster for each cell closest to the wall along $x$-direction for both upper and lower walls. The average of local height of these structures give the aggregation height, $h$.

Aggregation length and height are further averaged over 200 such steady state configurations.

\end{document}